\documentclass[%
 reprint, longbibliography,
 amsmath,amssymb,
 aps,
 prx,
]{revtex4-2}

\usepackage{graphicx}
\usepackage{dcolumn}
\usepackage{bm}
\usepackage[colorlinks=true,linkcolor=blue]{hyperref}
\usepackage{natbib}

\begin{document}

\preprint{APS/123-QED}

\title{Modelling nanomagnet vertex dynamics through Coulomb charges}

\author{Samuel D. Slöetjes}
\author{Matías P. Grassi}
\author{Vassilios Kapaklis}
\affiliation{Department of Physics and Astronomy, Uppsala University, Box 516, SE-75120 Uppsala, Sweden}%

\begin{abstract}
We investigate the magnetization dynamics in nanomagnet vertices often found in artificial spin ices. Our analysis involves creating a simplified model that depicts edge magnetization using magnetic charges. We utilize the model to explore the energy landscape, its associated curvatures, and the fundamental modes. Our study uncovers specific magnonic regimes and transitions between magnetization states, marked by zero-modes, which can be understood within the framework of Landau theory. To verify our model, we compare it with micromagnetic simulations, demonstrating a noteworthy agreement.
\end{abstract}

\maketitle

\section{Introduction}
Artificial spin ices (ASIs) have been widely utilized to study and explore intriguing physical phenomena such as frustration and magnetic monopoles \cite{Nisoli:2017hg, Rougemaille:2019ef,Advances_ASI}. These systems consist of large collections of periodically ordered nanomagnets -- {\it mesospins} -- which are coupled through magnetostatic interactions. The critical interactions in these artificial lattices occur at the vertices, and under certain conditions, these interactions can give rise to frustration \cite{morrison2013unhappy, Perrin:2016hj, Ostman:2018cp}. Because of their design flexibility and tunable complexity, these systems have garnered interest for use as magnonic crystals, where the freedom of placement of nanomagnets and their susceptibility to magnetic fields yield reconfigurable magnon bands \cite{Gliga:2013jw, Iacocca:2016gb, alatteili2023g}.

In simplified spin ice models, the building blocks are assumed to feature a rigid, single domain magnetization, which has been proven to be successful for predicting static configurations of the magnetization\cite{jensen2022flatspin}. Several numerical studies suggest that additional bending of the magnetization texture occurs at smaller length scales, which can provide emergent monopole-like excitations with a chiral character \cite{rougemaille2013chiral}, and affect the GHz magnetization dynamics \cite{Arora2022, lendinez2023, dion2023ultrastrong}. Moreover, we recently showed that texture bending at the edges can give rise to magnon driven thermal fluctuations between different metastable states in single nanomagnets \cite{sloetjes2021effect} and two coupled nanomagnets  \cite{sloetjes2022texture}. We also reported experimental signatures of such fluctuations in ac susceptibility and hysteresis measurements \cite{skovdal2023thermal}. More complex texture formation in ASI mesospins was observed to form as a result of intra-magnet interactions, and coupling to neighboring mesospins  \cite{saccone2023vertices,sasaki2022formation}. Moreover, a combination of single domain textures and flux closure textures has been utilized for realizing a reservoir computing hardware schemes with ASIs \cite{gartside2022reconfigurable}.

\begin{figure}
    \centering
    \includegraphics[width=0.6\columnwidth]{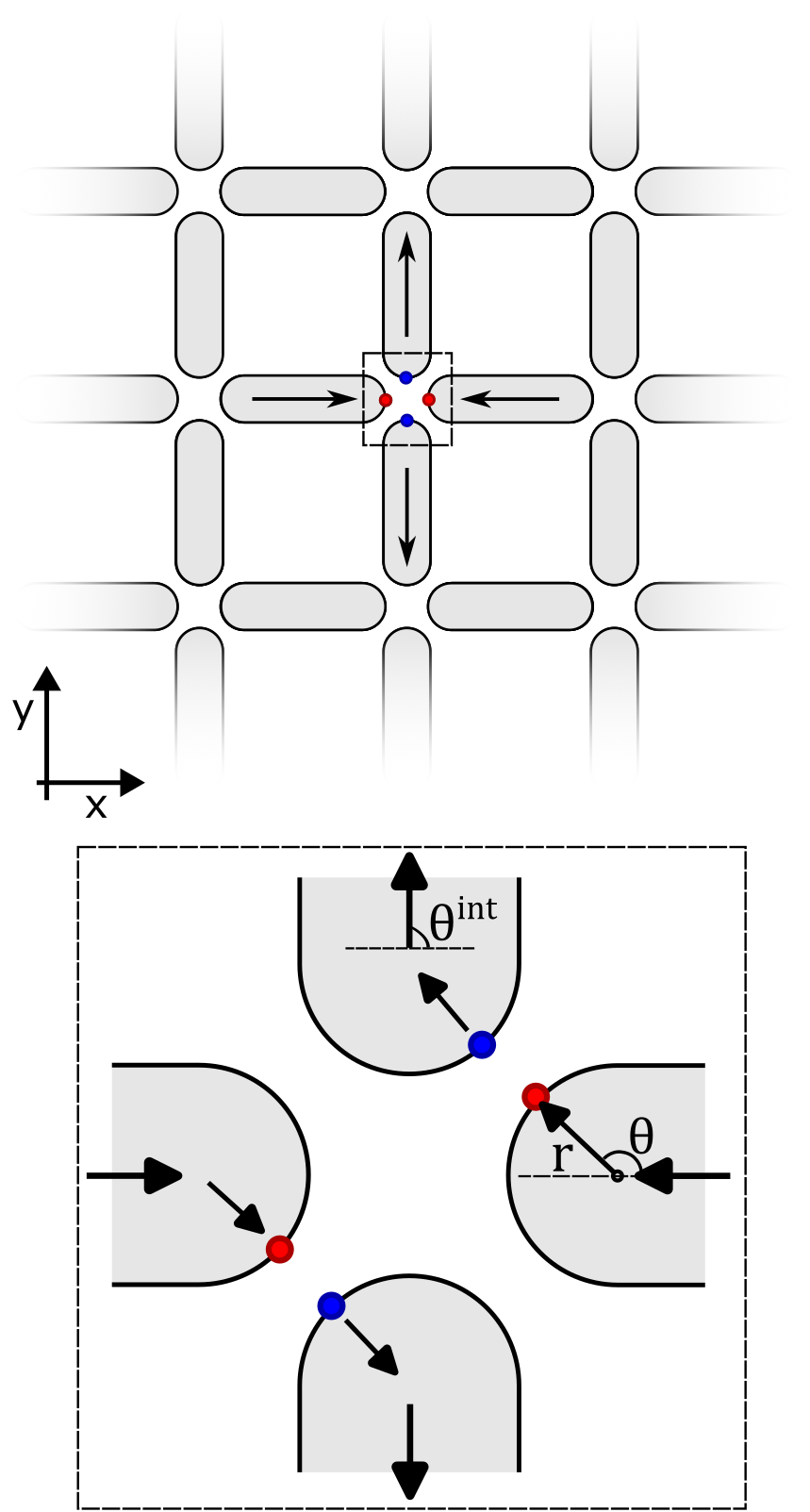}
    \caption{The mesospin magnetization texture in a Type I vertex is approximated in terms of interacting monopoles (red and blue) that are exchange coupled to the inner magnetization (black). }
    \label{fig:dip}
\end{figure}

These findings suggest a replacement of a rigid spin minimal model when simulating large ensembles of nanomagnets, namely one in which bending of the inner magnetization texture is also taken into account. 
Here, we make a first step toward this end, by proposing a simple model based on interacting Coulomb charges, reminiscent of the dumbbell model. We perform a systematic study of the static and dynamic phenomena that occur at the vertices as a result of the nanomagnets' internal excitations at the edges. We test this model for the four different types of vertices (Types I-IV), and for each vertex we identify the minima in the energy landscape, and calculate the static and dynamic modes as a function of the balance between exchange and dipolar energies. We find several interesting spectral features, such as mode crossing and mode softening to zero frequency as a function of the exchange stiffness. Finally, we utilize micromagnetic simulations to investigate the modes and find qualitative agreement between the two methods.

\subsection{Model system}

In our model system, the relevant magnetization of the nanomagnets in the edges is reduced to a single surface magnetic charge, akin to a dumbbell model. In the dumbbell model, the dipole interaction is mediated through charges, which can be regarded as monopoles. As such, the dipole interaction in the Hamiltonian is represented by a Coulomb-like term:
\begin{equation}
    E_{\mathrm{dip}} = D\,\sum_{ij}\frac{Q_iQ_j}{r_{ij}}.
    \label{dipolar_eq}
\end{equation}
Here, $Q_i$ represents charge $i$, and can assume values of $+1$ and $-1$, $r_{ij}$ is the distance between surface magnetic charges $i$ and $j$, and $D$ is the prefactor that sets the strength of the dipole interaction. The charges are bound to the edges of the nanomagnets, which are given by semicircles of radius $r$, see Fig. \ref{fig:dip}. A negative charge indicates the magnetization pointing into the structure, while a positive charge corresponds to the magnetization pointing out of the structure. As such, the different vertex types can be encoded into configurations of charges located at the edges of the nanomagnets.
Since the opposite charge of the nanomagnet is situated on a neighbouring vertex, it is sufficiently far removed that we neglect its effects on the dynamics of the vertex under consideration. 

The exchange interaction is given by:
\begin{equation}
    E_{\mathrm{ex}} =-A \sum_i cos(\theta_i - \theta_i^{\mathrm{int}}),
    \label{exchange_eq}
\end{equation}
\noindent
where $\theta_i$ is the in-plane rotation angle of the vector that from the center of the semicircle to the charge on the edge (see Fig. \ref{fig:dip}), and $\theta_i^{int}$ represents the angle of the inner magnetization of the magnet $i$ with the x-axis. The strength of the exchange stiffness is tuned through the parameter $A$. 

The total energy is then simply given as a sum of the exchange and dipole contributions:
\begin{equation}
    E = E_{\mathrm{ex}} + E_{\mathrm{dip}}. 
\end{equation}

A similar approach has been applied to investigate edge mode dynamics by \citet{mcmichael2006edge}, although in this case a macrospin was used to model the edge magnetization, instead of a single charge.
We set the dipole interaction strength equal to unity, i.e. $D=1$, and fix the radius of the semicircle, $r$, that defines the edge of the mesospin. As such, the only free parameter is the exchange stiffness, $A$, which can be expressed relative to $D$ using the ratio $A/D$. In ASI systems, this ratio is system specific and the most straightforward way to tune it is through the spacing between the nanomagnets, as smaller spacing increases the dipole coupling, decreasing $A/D$. Alternatively, $A$ can be tuned through choice of material. 

Since there are four mesospins involved in a vertex (see Fig. \ref{fig:dip}), the energy landscape, $E(\theta_1,\theta_2,\theta_3,\theta_4)$ is four-dimensional (4D), and each angle is varied from $\theta_i = -\pi/2$ to $\pi/2$ in 40 steps, such that $\Delta\theta=\pi/40$, resulting in a $40\times40\times40\times40$ matrix describing the energy landscape. We identify the lowest metastable states as a function of $A/D$ by simply finding the minima in the energy landscape minimizing the energy $E(\theta_1,\theta_2,\theta_3,\theta_4)$. Following \citet{smit1955ferromagnetic}, we determine the dynamic properties for each vertex by approximating the metastable minima as harmonic potentials, i.e., by assuming a paraboloid shape of the minima. This method has recently also been used to calculate the dynamics of skyrmions \cite{desplat2023eigenmodes}, and allows us to extract the curvatures and mode profiles as the eigenvalues and eigenvectors of the Hessian matrix, evaluated at the stable points. 

The Hessian matrix constitutes the matrix of second derivatives of the energy to the angle of rotation:

\begin{equation}
    \mathbf{H} = 
    \begin{bmatrix}
        E_{\theta_1,\theta_1} & E_{\theta_2,\theta_1} & E_{\theta_3,\theta_1} & E_{\theta_4,\theta_1}\\
        E_{\theta_1,\theta_2} & E_{\theta_2,\theta_2} & E_{\theta_3,\theta_2} & E_{\theta_4,\theta_2}\\
        E_{\theta_1,\theta_3} & E_{\theta_2,\theta_3} & E_{\theta_3,\theta_3} & E_{\theta_4,\theta_3}\\
        E_{\theta_1,\theta_4} & E_{\theta_2,\theta_4} & E_{\theta_3,\theta_4} & E_{\theta_4,\theta_4}
    \end{bmatrix}.
\end{equation}
\noindent
Here, we use the shorthand notation $E_{\theta_i,\theta_j}=\frac{\partial^2E}{\partial\theta_i\partial\theta_j}$. Before calculating this matrix, we recalculate the energy landscape with an increased resolution of $\Delta\theta=\pi/600\approx0.005$ around the minima, to ensure a well-defined curvature. Furthermore, care must be taken when numerically evaluating the Hessian matrix, therefore we use the seven-point stencil method to calculate the derivatives \footnote{the seven-point stencil method is numerically evaluated as:
\begin{equation}
\begin{split}
 \frac{df}{dx}=1/(60\Delta\theta)[-f(x-3\Delta\theta)+9f(x-2\Delta\theta)
 \\
 -45f(x-\Delta\theta) +45f(x+\Delta\theta)
 \\
 -9f(x+2\Delta\theta)+f(x+3\Delta\theta)]   
\end{split}
\end{equation}}.
The Hessian matrix is evaluated at the global minima of the energy landscape, which correspond to the stable states in the vertex. The eigenvalues, $\kappa_1, ..., \kappa_4$, and eigenvectors $\mathbf{v}_1, ..., \mathbf{v}_4$ of the Hessian matrix are then  evaluated by solving:
\begin{equation}
    \mathbf{H}\mathbf{v}_i=\kappa_i\mathbf{v}_i.
\end{equation}

The eigenvalues, $\kappa_i$, represent the curvatures in the energy landscape, and can be related to the mode frequency, $f_i$, via $f_i\propto \sqrt{\kappa_i}$. To make this relation more intuitive, it is instructive to consider the Kittel relation, given by $f=\mu_0 \gamma\sqrt{H_0}\sqrt{H_0 + M_{\mathrm{S}}}$, which is essentially the product of the curvature of the energy landscape in the in-plane and out-of plane directions, i.e. $f \propto \sqrt{\kappa_{\parallel}}\sqrt{\kappa_{\perp}}$ (see Eq. S2 in \citet{desplat2020entropy}). In our case, we choose to ignore the effect of the variation of the out-of-plane contribution, since we have seen previously that the precessional motion of the edge modes is dominated by the in-plane component \cite{sloetjes2021effect}. Moreover, the curvature in the energy landscape associated with pointing the magnetization out of the plane hardly changes upon variation of micromagnetic parameters (see Appendix B of \citet{sloetjes2022texture}). 

The corresponding spatial information of the dynamic modes can be found via the eigenvectors. For example, in the Type IV vertex, two normalized eigenvectors are:
\begin{equation}
\mathbf{v}_1 = 
    \begin{pmatrix}
        1\\
        -1\\
        1\\
        -1
    \end{pmatrix},
\mathbf{v}_4 = 
    \begin{pmatrix}
        1\\
        1\\
        1\\
        1
    \end{pmatrix},
\end{equation}
\noindent
here, $\mathbf{v}_1$ refers to a mode where neighbouring spins are oscillating out of phase, and $\mathbf{v}_4$ refers to a mode where all spins oscillate in phase. Thus, we interpret the sign and the magnitude of the vector components as the phase and the relative amplitude of the oscillation, respectively. 

It is noteworthy that this model also has potential applications for analyzing systems of interacting charged colloids in bistable traps. Such artificial systems have been studied theoretically and experimentally, and have displayed capabilities for hosting additional physics with respect to conventional artificial spin ice systems\cite{libal2006realizing,loehr2016defect,nisoli2018unexpected,ortiz2019colloquium}. For these colloidal model systems, only positively charged colloids were considered, however, in the case that negative charges are also added, our model could be directly mapped to these systems.

\section{Results}

For the static configurations, we inspect the absolute minima in the energy landscape under variation of $A/D$. Here, we can make a general observation applicable to all vertices, namely that in the limit $A/D \rightarrow\infty$ (exchange dominated) the magnetization curvature at the edges vanishes and the nanomagnet assumes a uniform magnetization. It is also worthy to note that in the opposite limit, $A/D \rightarrow 0$, the ground states corresponding to different Types are not identical, as the charge sign respects the energy minimization imposed by the inner magnetization, i.e., no domain wall is allowed. 

In the following section, we will inspect the dynamic modes for the Type I-IV  vertices (shown in Fig. \ref{fig:AllTypesDB}a) in the ASI system. The spectra are color coded according to their mode configurations \footnote{Usually spectral colors refer to the k-vector of a certain mode, however, this does not apply here since a k-vector is ill-defined in a four-spin system. To obtain a color for a spectral branch, we calculate a number, $N$, associated with the mode, which then selects a color from an RGB matrix. This number is calculated as 
\begin{equation}
    N = int[\lvert\mathbf{V}\cdot (2 \; 4 \; 6 \; 8) \rvert],
\end{equation}
where $\mathbf{V}$ represents the eigenvector associated with an eigenvalue that relates to a frequency. The absolute value ensures that degenerate modes give the same color, for example, it can be seen that $\mathbf{V} = (1 \; {-}1 \; 1 \; {-}1)$ gives the same color as $({-}1 \; 1 \; {-}1 \; 1)$. The $int$ command means rounding to the nearest integer, which is necessary for selecting one set of RGB values from a matrix, i.e. \texttt{RGBValues = RGBMatrix(N,:)}}. Afterwards we use micromagnetic simulations to corroborate our analytic model, and to study the system under thermal excitation. We will indicate the modes as a vector, where the elements represent the phase according to $($top right down left$)$ spins, i.e. starting at the upper element, and cycling clockwise through the configuration. For example $(++--)$, represents the top and right spins in phase, but in opposite phase with the lower and left spins.

\subsection{Results from charge model}

We vary the ratio of exchange stiffness to dipole contribution $A/D$, from 0 to 3 with a step of 0.005, and map out the resulting spectra for the different vertex Types found in ASI systems.

\subsubsection{Type I}

The Type I vertex has an $($out in out in$)$ configuration of spins. The resulting spectrum shown in Fig. \ref{fig:AllTypesDB}b shows that the highest frequency mode is one where all spins oscillate in unison, whereas the lowest frequency mode has the same symmetry as the vertex, $(+-+-)$, and goes to zero for a critical value of $A_c/D\approx1$. Below this value, the dominating dipole interaction causes the existence of two degenerate metastable states associated with opposite bending of the magnetization in the vertex (see inset of Fig. \ref{fig:AllTypesDB}b), while for $A/D>1$ strong exchange contribution causes the edge magnetization to align with the inner magnetization of the vertex. Thus, the zero mode signifies a transition, upon which a dynamic mode ``freezes in". The two metastable states below the transition break the double mirror symmetry of the vertex, which causes the $(++--)$ and the $(+--+)$ dynamic modes to have different frequencies. At the transition, these two modes can be seen to merge into a (nearly) single mode, which persists as $A/D\rightarrow\infty$.

\subsubsection{Type II}

\begin{figure}
    \centering
    \includegraphics[width=1\columnwidth]{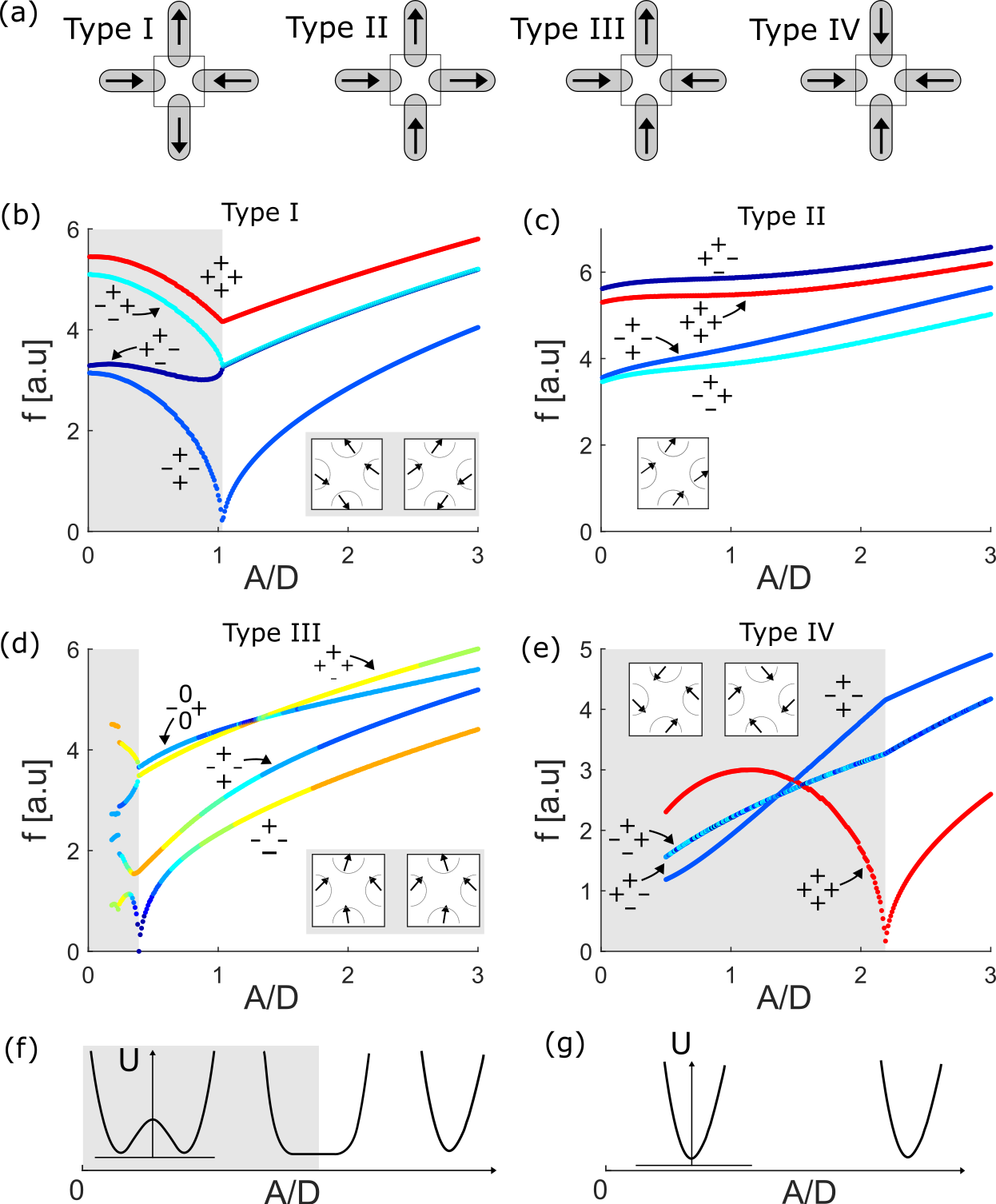}
    \caption{(a) Type I, (b) Type II, (c) Type III, and (d) Type IV edge spectra for different values of $A/D$. The gray shading indicates that the system is in the dipole dominated region, below the transitions, and the corresponding doubly degenerate groundstates are shown in the insets with gray background. Since the Type II vertex does not feature a transition, the gray shading is absent, and a single state is shown. (f) Sketch of the energy landscape at different values of $A/D$, for the Type I, III, and IV vertices, (g) vice versa for the Type II vertex}
    \label{fig:AllTypesDB}
\end{figure}

The Type II vertex has an $($out out in in$)$ configuration of spins. Here, the four different modes are clearly split, where the $(+--+)$ has the highest frequency, and the $(++--)$ mode, which reflects the vertex symmetry, has the lowest frequency, see Fig. \ref{fig:AllTypesDB}c. The two modes in between are seen to approach one another as $A/D$ is increased. One striking observation is that there is no mode that freezes in, and as such there is no clear distinction between a dipole and exchange dominated regime. Instead, the moments in the inner part of the vertex slowly move from pointing diagonally in the same direction, to the direction of the Type II vertex (see Supplemental Material). This lack of a transition likely stems from the fact that there is only a single metastable minimum, even as $A/D\rightarrow0$. This arises from the symmetry of the Type II configuration, which allows for a single edge configuration (see inset of Fig. \ref{fig:AllTypesDB}c). Additionally, when comparing to the Type I spectrum, the spectrum does not feature any degeneracy of modes, which we attribute to the fact that this vertex possesses only a single mirror symmetry axis, across the diagonal.

\subsubsection{Type III}

The Type III vertex has a $($out in in in$)$ configuration of spins. The spectrum stands out in its colorfulness, see Fig. \ref{fig:AllTypesDB}d, which indicates that the branches change their mode character as a function of $A/D$, through continuous change of the amplitudes in each edge (cf. Supplemental Material). This relative difference in amplitudes for the different edges means essentially that the normal modes are not pointing in the diagonal directions of the 4D energy landscape. This lack of diagonal modes is fundamentally due to the fact that these normal modes are incommensurate with the symmetry of the Type III vertex, and as such the dynamics have to adept for each value of $A/D$.

It can be seen that the there is a transition from the dipole dominated to the exchange dominated regime at $A/D\approx0.3$, where the mode that freezes in indeed has a symmetry of $(+---)$, the same as the vertex configuration, where the amplitude of each oscillation in this mode varies throughout the sweep of $A/D$. The two metastable states that exist below the critical value of $A/D$ show a subtle difference, which is only pronounced in the upper and lower spins. The spectrum also features a mode where the upper and lower edges do not oscillate, but the right and left edges oscillate out of phase, i.e. $(0+0-)$.

\subsubsection{Type IV}

For the Type IV configuration, all spins are pointing into the vertex, which in our model is represented by four positive charges. Indeed, it can be seen that the mode that goes to zero frequency has the same symmetry as the vertex configuration, $(++++)$, see Fig. \ref{fig:AllTypesDB}e. Below the transition, the ground state is doubly degenerate, where each degenerate state can be seen to represent a vortex with a certain handedness. In this regime, a mode inversion can be observed at $A/D\approx1.5$. Throughout the whole spectrum, the $(++--)$ and $(+--+)$ modes are seen to overlap, which originates from the high amount of mirror symmetry axes at the vertex. Before the inversion, the $(++++)$ mode has the highest frequency, while above the transition, it is the $(+-+-)$ mode that has the highest frequency.

\subsection{Relation to Landau theory}
The transitions indicated by the zero-mode can be seen in the context of Landau theory, appropriate for describing second-order phase transitions \cite{landau2013electrodynamics}. The free energy is approximated to the first two even terms of a Taylor expansion as:
\begin{equation}
    F(\phi, A/D) = F_0 + \alpha(A/D)\,\phi^2+\beta(A/D)\,\phi^4
\end{equation}

Here $\phi$ is the order parameter, and $\alpha(A/D)$ and $\beta(A/D)$ are prefactors that depend on the control parameter $A/D$ (which is usually given by the temperature), and $\beta>0$. If $\alpha>0$, the energy has a single minimum, which  means that the expectation value of the order parameter vanishes, $\langle \phi\rangle=0$. When this parameter is decreased to $\alpha=0$, the energy landscape at the minimum is flat, and the system is at the critical state. When $\alpha$ is further decreased to $\alpha<0$, the system obtains two degenerate minima with $\langle \phi\rangle\neq0$. In our case, the order parameter $\phi$ would be the deviation of the edge magnetization with respect to the inner magnetization. For Type I, III, and IV, the dipole dominated regions correspond to $\alpha>0$, and the system has two degenerate metastable minima, see Fig. \ref{fig:AllTypesDB}f. At the transition value for $A/D$, $\alpha=0$, which leads to a zero-mode in the spectrum, as the flatness essentially means that one of the eigenvalues is zero, $\kappa=0$, and $f=0$. When $A/D$ is further increased into the exchange dominated region, $\alpha<0$, and the energy landscape has a single minimum, where again a positive curvature is obtained, such that $f>0$. For the Type II vertex, there is only one minimum throughout the range of $A/D$ values, such that $\alpha>0$  is always the case, see Fig. \ref{fig:AllTypesDB}g, and consequently the frequency always has a finite value, $f>0$.

\section{Micromagnetic simulations}

We performed micromagnetic simulations to corroborate the results from our minimal model. For this, we used \textsc{MuMax}$^3$ \cite{vansteenkiste2014design} to simulate an ASI vertex, consisting of stadium shaped islands with a length, width, and thickness of 450 nm, 150 nm, and 1 nm, respectively. The spacing between the nanomagnets in both the vertical and horizontal directions is 100 nm. The material properties are based on $\delta$-doped Pd(Fe) \cite{papaioannou2010}, which is given by a saturation magnetization $M_s = 1700$ kA/m, and the damping is set to $\alpha=0.001$. We vary the exchange stiffness from $A_{\mathrm{ex}}=1\times10^{-11}$ J/m to $A_{\mathrm{ex}}=1\times10^{-10}$ J/m in steps of $1\times10^{-12}$ J/m. The cell size is 2.5 nm in the in-plane directions and 1 nm in the out-of-plane direction.

The spectra are generated using the ringdown method \cite{mcmichael2005magnetic}, which means that the magnetization is first relaxed in a small displacement field, $\mathbf{H}_D$, which is consequently removed, after which the magnetization rings down during 10 ns. The corresponding timetraces of the magnetization components, $m_x$, $m_x$, and $m_z$ are then recorded and finally Fourier transformed to obtain the spectral response:
\begin{equation}
    m(\omega) = \int_0^{t_{\text{end}}} m(t)e^{i\omega t}dt,
\end{equation}
where $\omega=2\pi f$ and $t_{\text{end}}=10$ ns.

The modes are locally excited at the edges of the nanomagnets, and the excitation is in the out of plane direction, i.e. $\mathbf{H}_D=(0,0,H_D)$, to ensure that the excitation field is always perpendicular to the static magnetization. Distinguishing between different modes in the spectra happens by encoding the mode in question into the excitation field. As such, each edge receives a field $\mathbf{H}_D$ which is pointing in either $+z$ or $-z$ directions for the different edges, depending on the phase of the oscillation that is being probed. For example, to excite the $(+-+-)$ mode, the field applied to edge 1 is $\mathbf{H}_D=(0,0,+H_D)$, edge 2 has $(0,0,-H_D)$, edge 3 has $(0,0,+H_D)$ and edge 4 has $(0,0,-H_D)$. From this, the phase distributions of the different branches in the spectra can easily be derived. The spectra with all the separate excitation modes can be found in the Supplemental Material, while in Fig.~\ref{fig:micromagSpec} these spectra are combined, i.e.:

\begin{equation}
\begin{split}
    m^{total}(f) = m^{++++}(f) + m^{+-+-}(f) \\+m^{++--}(f) + m^{+--+}(f)
\end{split}
\end{equation}

\begin{figure}
    \centering
    \includegraphics[width=1\columnwidth]{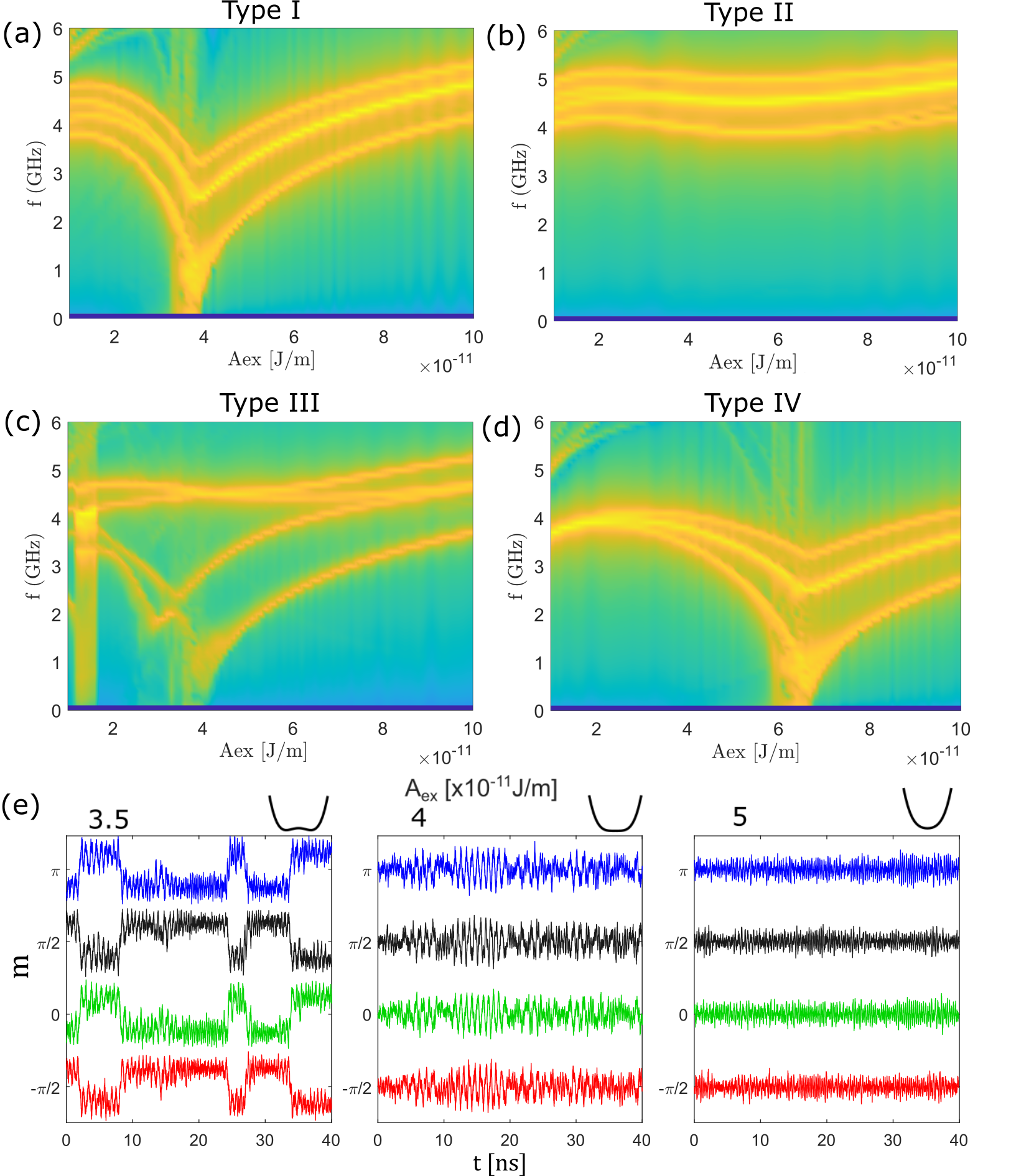}
    \caption{(a) Type I, (b) Type II, (c) Type III, and (d) Type IV edge spectra for different values of $A/D$. (e) Timetraces of the transverse moment in the Type I vertex, for different values of the exchange stiffness, indicating the second-order transition. The blue, black, green, and red lines correspond to the right, top, left, and bottom nanomagnet, respectively. The different energy landscapes are shown on the top of the graphs.}
    \label{fig:micromagSpec}
\end{figure} 

It can be seen that all spectra qualitatively agree with those generated with the charge model, also in terms of the modes associated with the branches. The spectra for Type I, III, and IV all feature a mode that goes to zero, whose spatial arrangement is given by the symmetry of the vertex (see Supplemental Material). These zero modes are seen to be accompanied by peaks extending over a large range of frequencies, indicated by the low intensity vertical stripes in the spectrum. Since critical phenomena are characterized by modes at all frequencies, we interpret these vertical stripes as phenomena related to criticality. For the Type I, the mode splitting can indeed be observed for $A_{ex}<3.8\times10^{-11}$J/m. In can be seen in the Supplemental Material that the order of the modes is the same as in the charge model. Small deviations between the two models can be seen, for example, in the bending of the $(+--+)$-mode below the transition. We ascribe this to the effect of simulating round structures in a micromagnetic framework, which leads to corrugated edges and induces a local anisotropy, thereby influencing the metastable positions and curvatures in the energy landscape. In the spectrum of the Type II vertex, indeed a zero mode is absent, and the $(+-+-)$ mode is seen to approach the $++++$ mode as $A_{\mathrm{ex}}$ increases. The Type III spectrum (Fig. \ref{fig:micromagSpec}c) shows the same peculiar behaviour around the transition as the results from the charge model. Moreover, the crossing of the upper two branches is reproduced at $A_{\mathrm{ex}}\approx4.2\times10^{11}$ J/m. The Type IV spectrum (Fig. \ref{fig:micromagSpec}d) features three branches, where the highest intensity is seen for the middle branch, which corresponds to the degenerate $(++--)$ and $(+--+)$ modes. The mode that goes to zero is indeed the $(++++)$ mode. The only part of the spectrum that is not reproduced in the Hessian model is for low $A_{\mathrm{ex}}$, which in the charge model features an inversion of modes, while in micromagnetics is seen to converge to a single mode. We ascribe this to the softening of the inner texture of the nanomagnets, which is not taken into account in the Hessian model, and can have an impact on the magnetization dynamics. A detailed summary of the results is shown in the Supplemental Material.

In order to emphasize the magnonic effects on the thermal behaviour of the vertex magnetization, we added a thermal excitation to the Type I vertex, in the form of a fluctuating field \cite{Leliaert_thermal_mumax3}, for three different values of the exchange stiffness around the transition, as shown in Fig. \ref{fig:micromagSpec}e. This thermal excitation is expected to increase the linewidth of the spectra, as was observed in a previous work\cite{sloetjes2021effect}. Below the transition value $A_{ex}<A_C$, the system fluctuates between two metastable states, corresponding to a bistable energy landscape. At the transition, $A_{ex}=A_C$, the system response features a complex oscillatory pattern with many frequency components, as expected in a critical regime, indicating a flat energy landscape. Above this value ($A_{ex}>A_C$), these fluctuations vanish, and a single, well-defined mode is observed.

Artificial spin systems with alternatively shaped building blocks have recently also been introduced, for example a needle shaped nanomagnet. In the case of a needle shaped nanomagnet, we expect extra anisotropy to be present, favoring collinear alignment with the nanomagnet interior, thus lowering the transition value for $A/D$. In realistic systems, the present of roughness along the nanomagnet edges will most likely form pinning centers, adding small potential wells to the energy landscape.

It is worth to note that we have also investigated a minimal model in which the magnetization in the edge was represented by a single magnetic dipole, which is exchange coupled to the interior magnetization. However, this model did not show agreement with the \textsc{MuMax}$^3$ simulations. More details on the results of this model can be found in the Supplemental Material.

\section{Conclusion}

In order to study the edge magnetization in the ASI vertex, we have introduced a simple model that approximates the mesospin edge magnetization as a magnetic charge on the mesospin semicircle boundary, which is further exchange coupled to the inner mesospin magnetization. The dynamics were calculated with a Hessian matrix, revealing transitions between different regimes, signified by a mode that goes to zero frequency, and reflects the symmetry of the respective vertex type. In addition, we analyzed the results in the context of Landau theory, and found that the transitions for the Type I, III, and IV bear similarities with second-order transitions. Finally, the model was tested against micromagnetic simulations, and showed qualitative agreement. 

Apart from being accurate and easy to implement, this Coulomb-charge model offers an immediate insight into the underlying physics behind the magnetization dynamics, due to the intuitive interpretation of the eigenvectors of the Hessian matrix. 

The charge model can also be applied to study the edge-magnetization dynamics of extended metamaterial lattices featuring complex arrangements, lowering the demand on computation power needed to understand the magnetization dynamics in spatially extended mesospin lattices.  

\section{Acknowledgements}
\noindent
We wish to thank Prof. Bj\"orgvin Hj\"orvarsson for fruitful discussions. S.D.S. and V.K. acknowledge support from the Swedish Research Council (Project No. 2019-03581). M.P.G. and V.K. also acknowledge support from the Carl Trygger Foundation (Project No. CTS21:1219).
\\
\\
\noindent
The authors have no conflicts of interest to disclose



%

\pagebreak
\onecolumngrid
\newpage
\begin{center}
\textbf{\large Supplemental Material: Modelling nanomagnet vertex dynamics through Coulomb charges}
\end{center}
\setcounter{equation}{0}
\setcounter{figure}{0}
\setcounter{table}{0}
\setcounter{page}{1}
\makeatletter
\renewcommand{\theequation}{S\arabic{equation}}
\renewcommand{\figurename}{Supplementary FIG.}
\renewcommand{\thefigure}{{\bf \arabic{figure}}}
\renewcommand{\bibnumfmt}[1]{[S#1]}
\renewcommand{\citenumfont}[1]{S#1}
\renewcommand{\thepage}{S-\arabic{page}}

\section{Static magnetization in the charge model versus micromagnetic model}

\begin{figure}[htb]
    \centering
    \includegraphics[width=1\columnwidth]{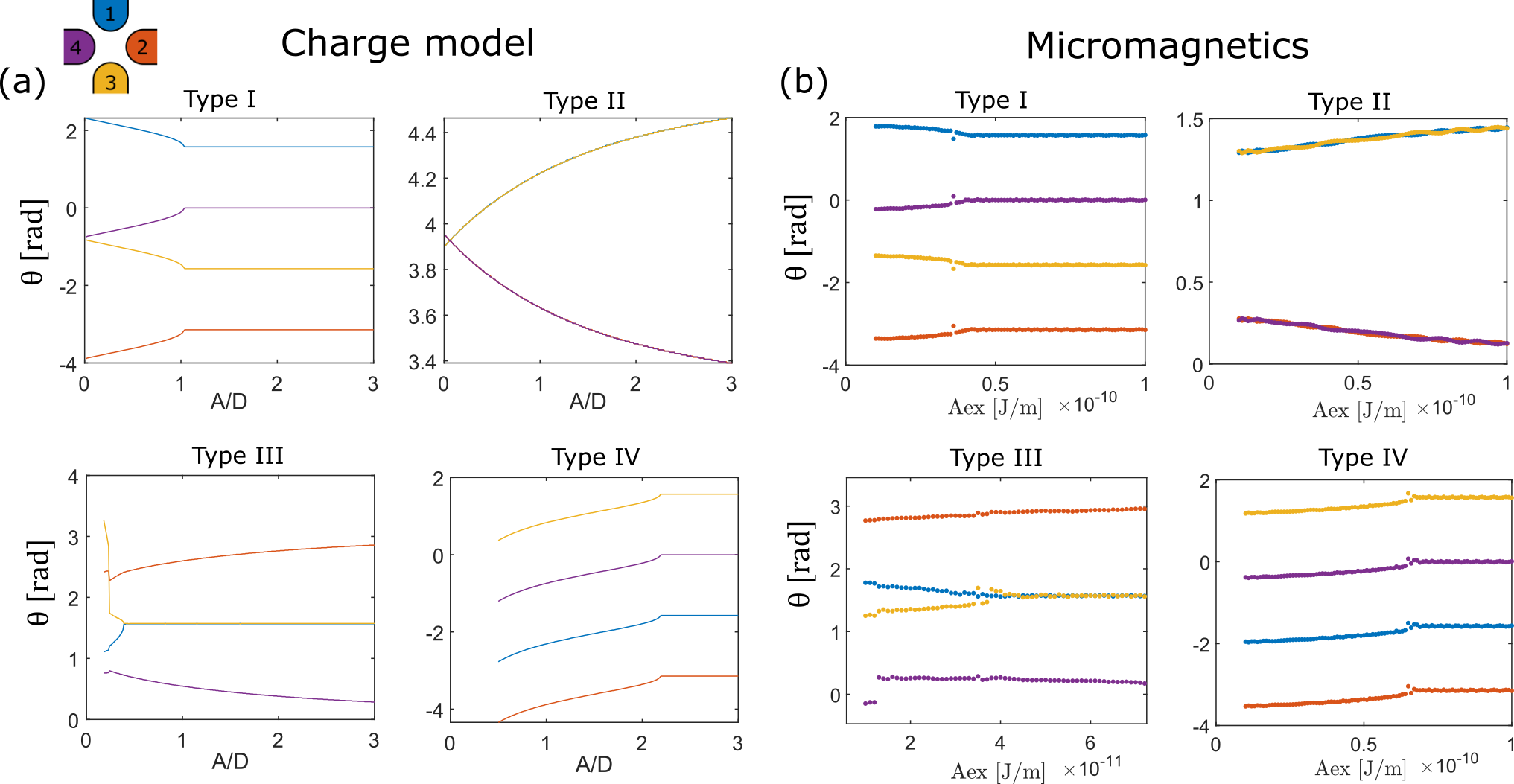}
    \caption{Static magnetizations for the four different types, obtained using the charge model. The colors refer to the edges as shown in the legend in the top left corner.}
    \label{fig:statdb}
\end{figure}

We calculated the magnetization groundstates in terms of the angle of the edge magnetization with the x-axis, $\theta$, for different values of $A/D$. We did this for the four different types, using both the charge model and the micromagnetic model, see Fig. \ref{fig:statdb}a and b, respectively. Good agreement can be seen between the two models, especially for the Type I and Type IV vertices. For the Type II configuration, the charge model gives more bending at low $A/D$, the same can be seen for Type III. 

\newpage
\newpage
\section{Eigenvectors}

\begin{figure}[htb]
    \centering
    \includegraphics[width=1\columnwidth]{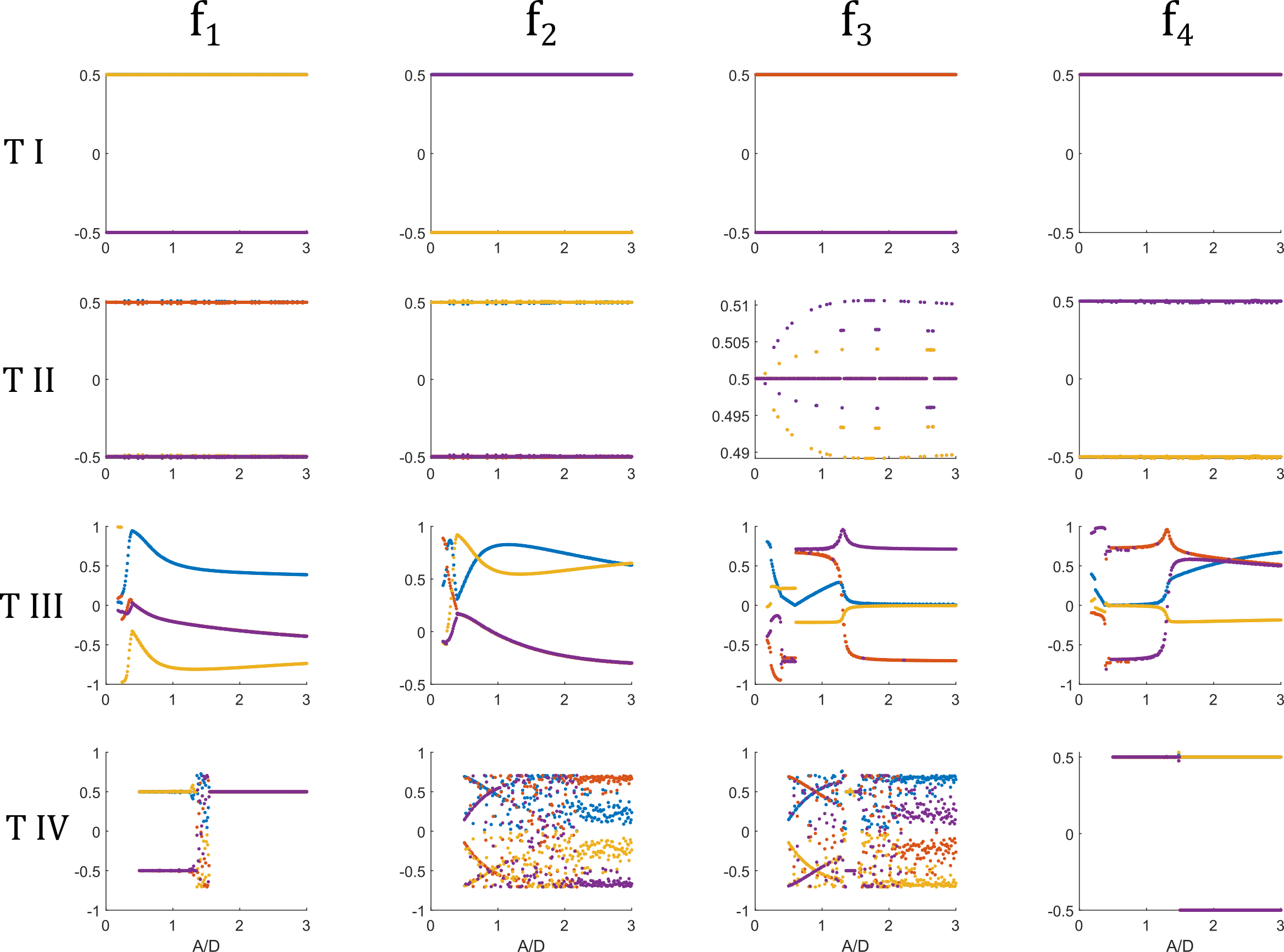}
    \caption{Eigenvectors for the different types and modes. The types are ordered from top to bottom, and the frequencies from left to right, in order of increasing frequency. The line colors indicate the edges, as shown in the legend in the upper left corner.}
    \label{fig:eigenvectors}
\end{figure}

The eigenvectors are plotted in Fig. \ref{fig:eigenvectors}, for different vertex types and different eigenfrequencies. Each line represents one edge, and the absolute value is the amplitude of the oscillation. For these plots, similar modes with opposite signs (for example $(+-+-)$ and $(-+-+)$) were reduced to one mode (i.e. $(-+-+)$ $\rightarrow$ $(+-+-)$) by always multiplying by $-1$ if the first element is negative.

It can be seen that Types I, II, and Type IV after the transition have a constant oscillation throughout the sweep of $A/D$. Type III stands out for its high variation in amplitude of the different edges.

\newpage

\section{Results for micromagnetic simulations for each excitation symmetry}

\begin{figure}[p]
    \centering
    \includegraphics[width=1\columnwidth]{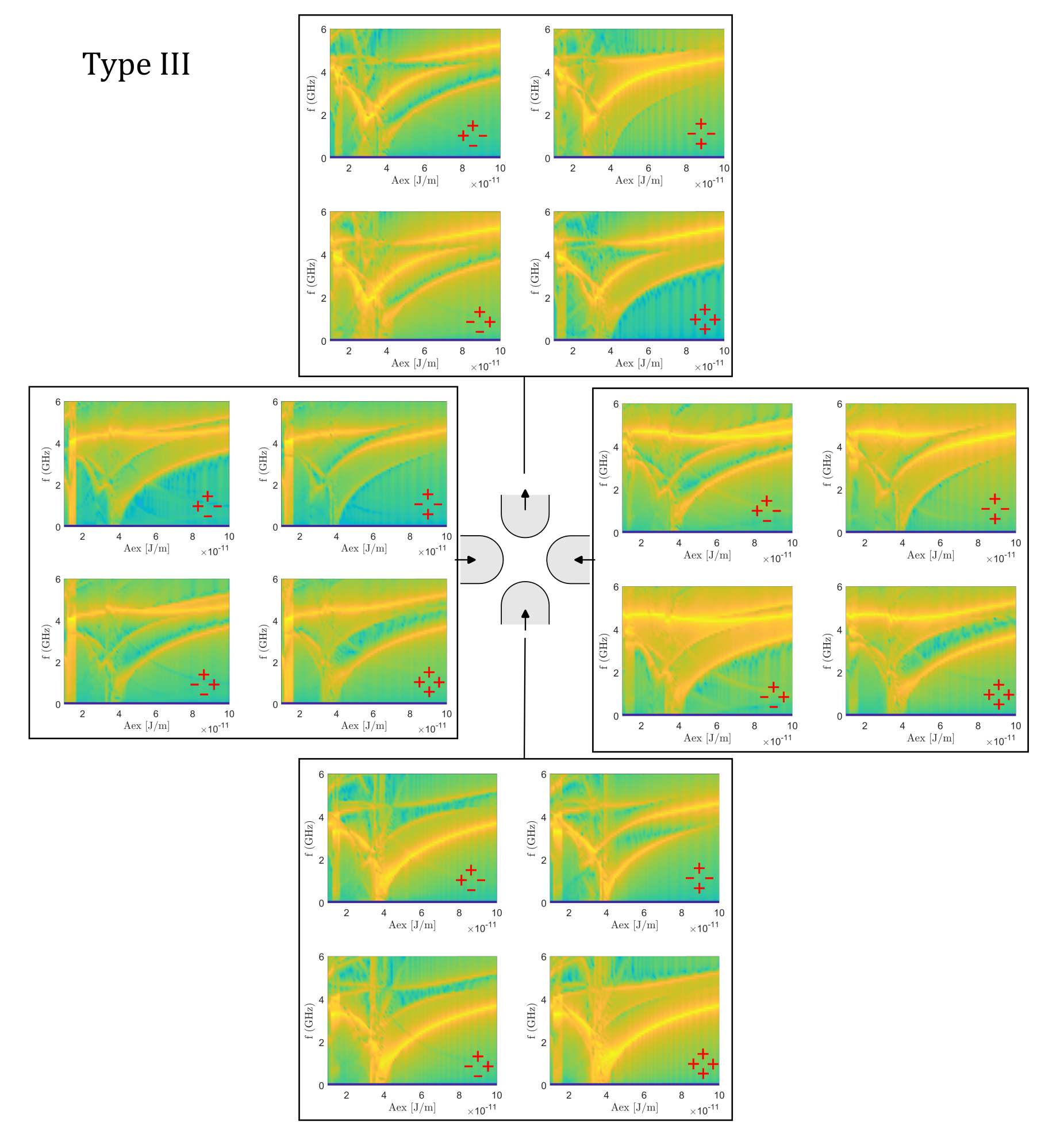}
    \caption{Spectra for the edge magnetization in Type III, recorded for each edge, with different excitation symmetries per edge.}
    \label{fig:MuSpec3}
\end{figure}

In this section, the separate spectra are shown that were generated in MuMax. The spectra shown in the main manuscript are the superposition of the separate spectra in this section. For Type I, II and IV, the modes were distributed equally over the four edges in the vertex, therefore we show the spectral response for just a single edge. For the Type III vertex, the modes were highly inhomogeneously distributed in the vertex, therefore we show the response separately for every edge (Fig. \ref{fig:MuSpec3}). In each figure, the field excitation symmetries are shown in red.

\begin{figure}[htb]
    \centering
    \includegraphics[width=0.9\columnwidth]{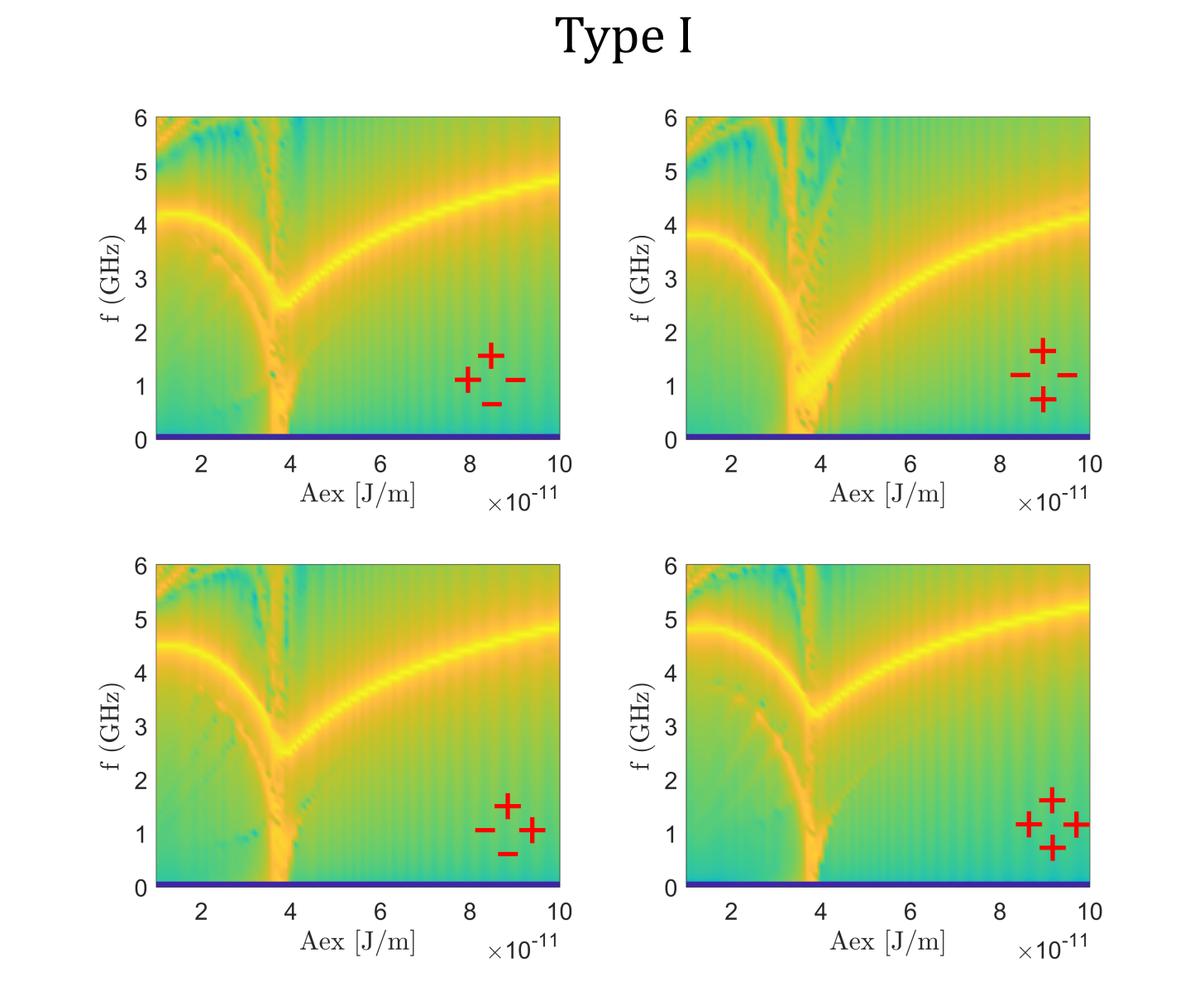}
    \caption{Spectra for the edge magnetization in Type I, with different excitation symmetries.}
    \label{fig:MuSpec1}
\end{figure}

\begin{figure}[p]
    \centering
    \includegraphics[width=0.9\columnwidth]{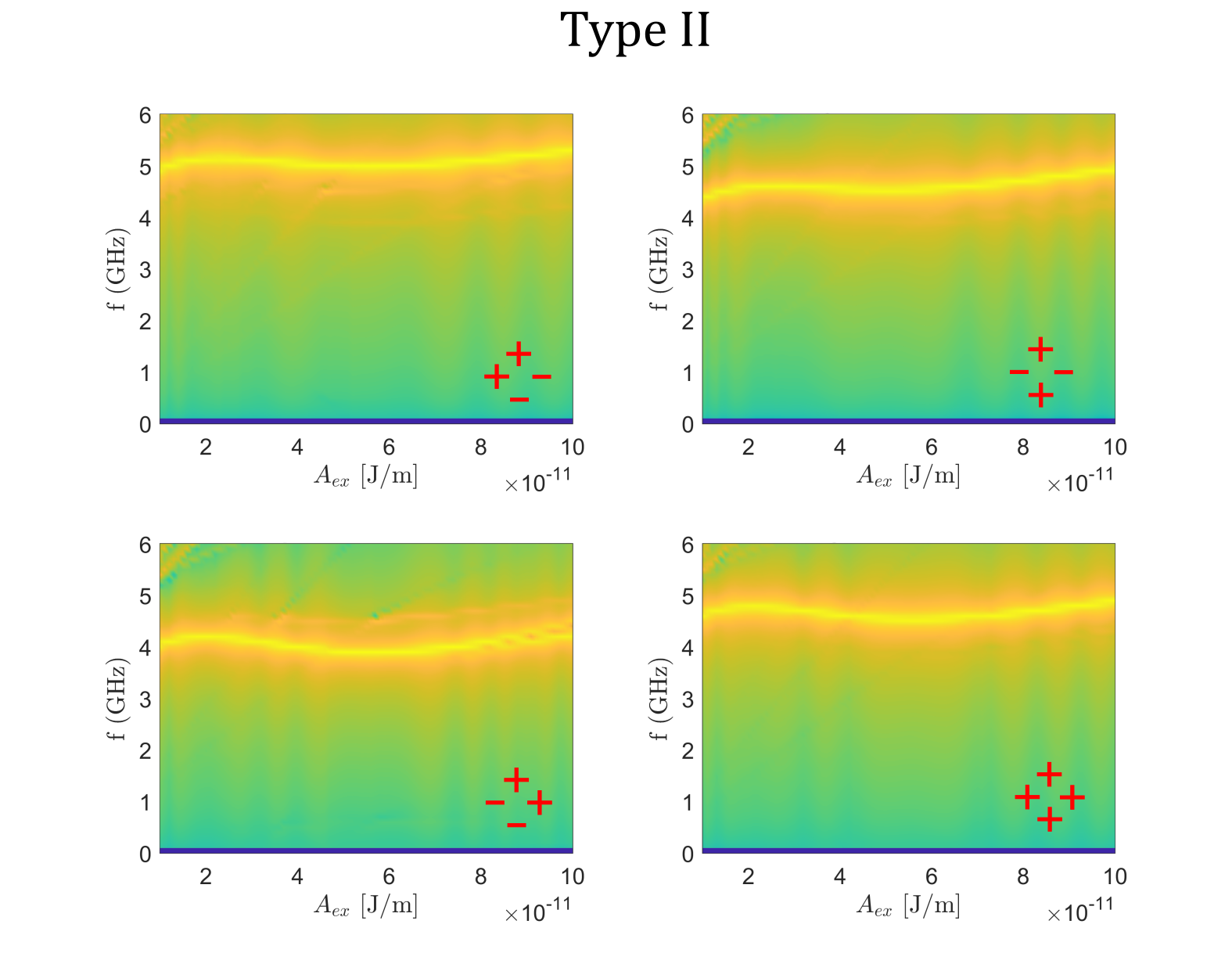}
    \caption{Spectra for the edge magnetization in Type II, with different excitation symmetries.}
    \label{fig:MuSpec2}
\end{figure}

\begin{figure}[p]
    \centering
    \includegraphics[width=0.9\columnwidth]{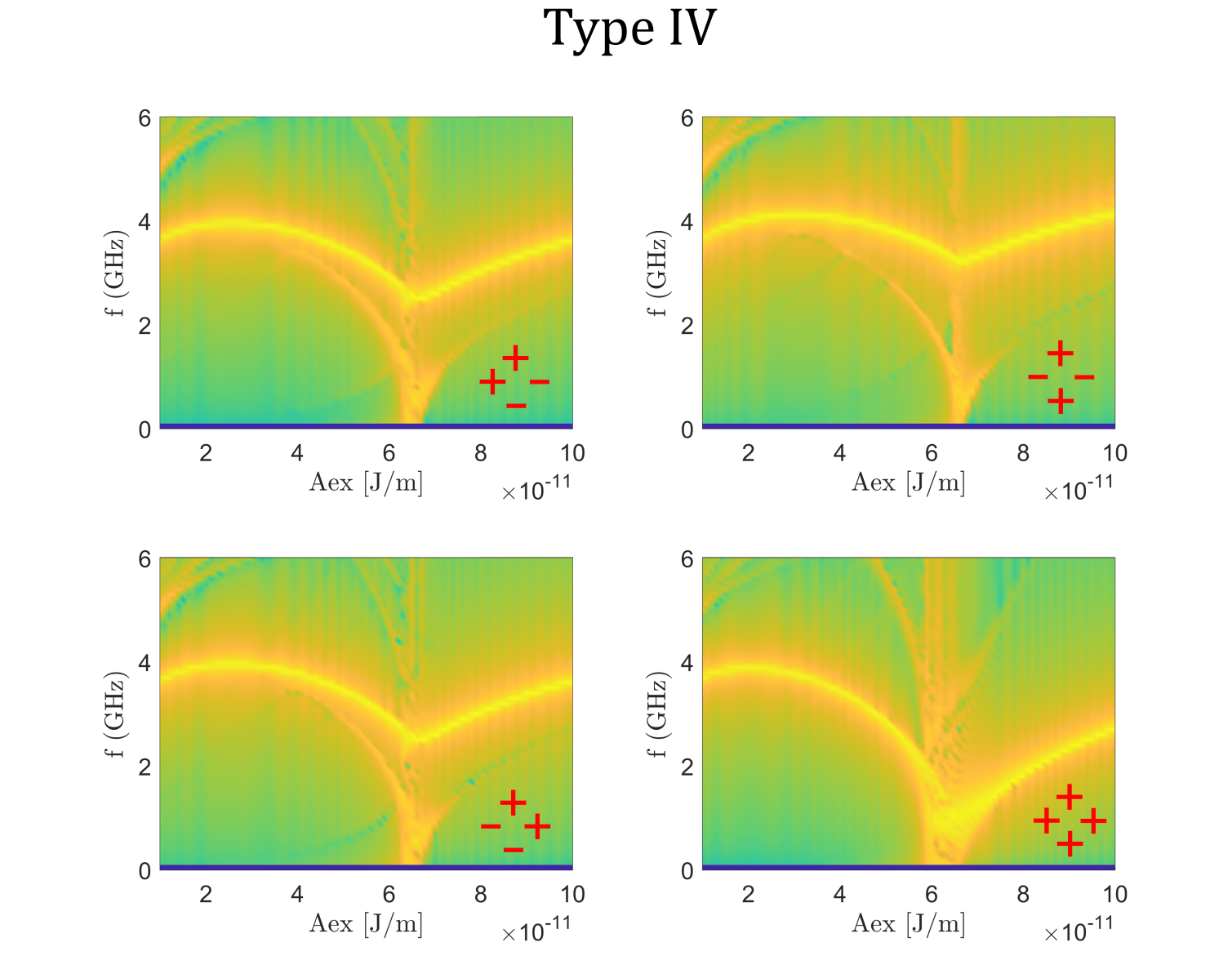}
    \caption{Spectra for the edge magnetization in Type IV, with different excitation symmetries.}
    \label{fig:MuSpec4}
\end{figure}

\section{Macrospin model results}

In addition to the charge model, we have explored the possibility of modelling the edge magnetization by a single macrospin (or magnetic dipole) at the center of the edge. 

Within this model, like in the charge model, the total energy is given as a sum of the exchange and dipolar contributions:

\begin{equation}
    E = E_{\mathrm{ex}} + E_{\mathrm{dip}}.
\end{equation}

However, the dipole interaction is given by

\begin{equation}
    E_{\mathrm{dip}} = -D\sum_{i\neq j}\frac{1}{r_{ij}^3}[3(\mathbf{m}_i\cdot\mathbf{\hat{r}})(\mathbf{m}_j\cdot\mathbf{\hat{r}}) - \mathbf{m}_i\cdot\mathbf{m}_j],
    \label{dipolar_eq}
\end{equation}

and the exchange interaction is given by 

\begin{equation}
    E_{\mathrm{ex}} = -A\sum_i \mathbf{m}_{i}^{\mathrm{int}}\cdot\mathbf{m}_{i} = -A \sum_i cos(\theta_i - \theta_i^{\mathrm{int}}),
    \label{exchange_eq}
\end{equation}
\noindent

where $r_{ij}$ is the distance between the magnetic moments $m_i$ and $m_j$, and $\theta_i$ and $\theta_i^{int}$ is the in-plane rotation angle of the $i$th edge- and interior spin with respect to the horizontal axis, respectively. As in the charge model, the strength of the dipole interaction and the exchange stiffness are given by $D$ and $A$, respectively.

\begin{figure}[p]
    \centering
    \includegraphics[width=0.9\columnwidth]{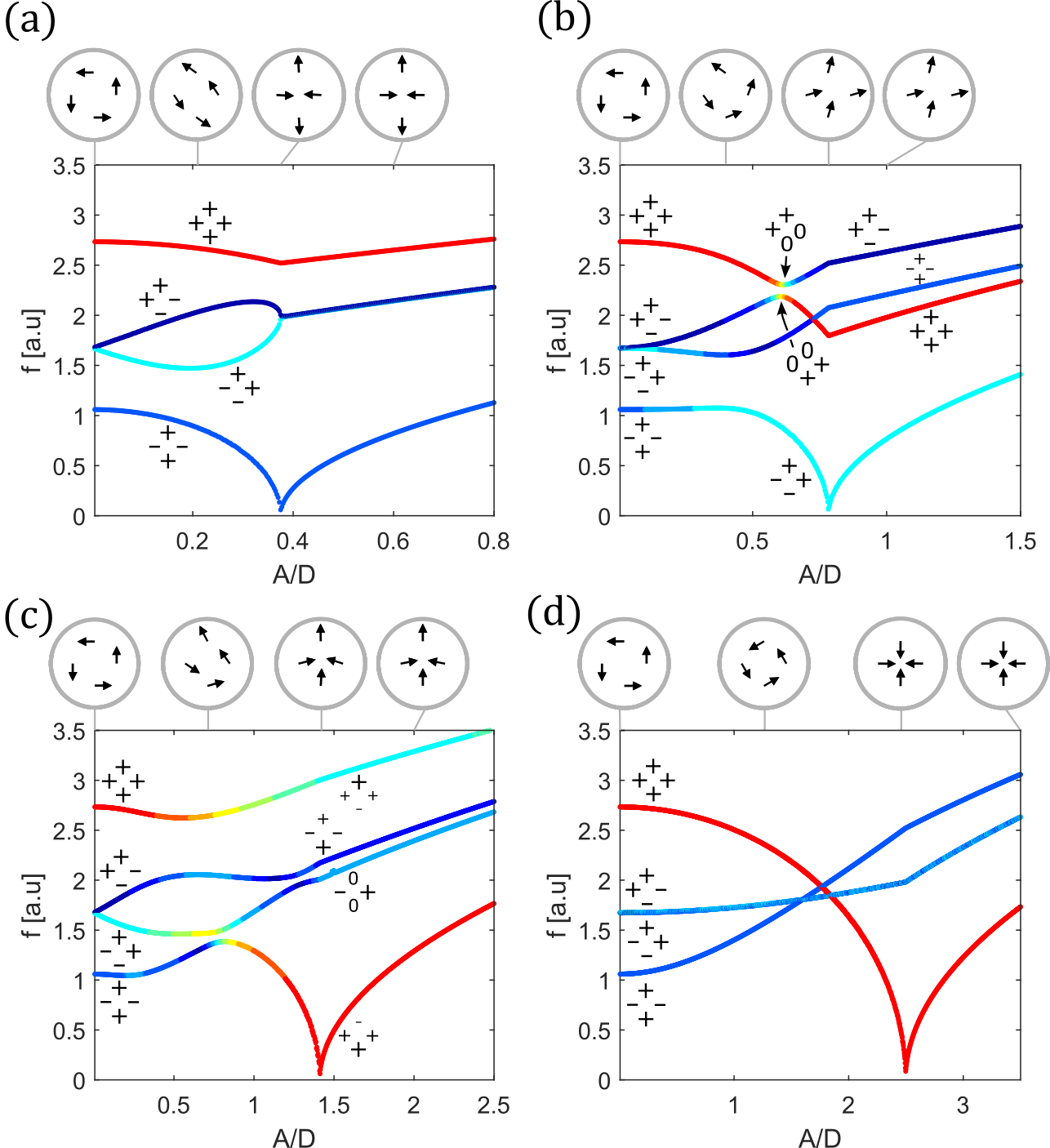}
    \caption{(a)Type I , (b) II , (c) III  and (d) IV  edge spectra obtained with the dipole model. The signs indicate the symmetry of the dynamics modes, while the arrows indicate the static configuration of the edge magnetization.}
    \label{fig:SM_dipole_model}
\end{figure}

Despite the similarities between the two models, the obtained spectra are qualitatively different, as shown in Fig. \ref{fig:SM_dipole_model}. The main difference with the charge model and the micromagnetic simulations is the presence of a zero-frequency mode and the subsequent transition observed for the Type II vertex (see Fig. \ref{fig:SM_dipole_model}b). These differences can be understood by the dipole consisting of both a + and a - charge, i.e. it has an uncompensated charge on the other end. This is in contrast to the charge model, which consists of single charges at the edges. Additionally, the dipole model lacks spatial mobility of the dipole, as it is always fixed at the center of the edge. Therefore, the distances $r_{ij}$ are solely defined by the vertex structure and cannot evolve over time.

\end{document}